\documentclass[12pt]{iopart}

\usepackage{graphicx}
\usepackage{amssymb}

\usepackage{epstopdf}

\begin{document}

\title{Monte Carlo Generation of Bohmian Trajectories}
\footnote{To be published in {\it J. Phys. A: Math. Theor.}}

\author{T M Coffey$^1$, R E Wyatt$^2$ and W C Schieve$^1$}

\address{$^1$ Department of Physics and Center for Complex Quantum Systems, 1 University Station C1600, University of Texas, Austin, TX 78712, USA}

\address{$^2$ Department of Chemistry and Biochemistry and Institute for Theoretical Chemistry, 1 University Station A5300, University of Texas, Austin, TX 78712, USA}

\ead{tcoffey@physics.utexas.edu}

\begin{abstract}
We report on a Monte Carlo method that generates one-dimensional trajectories for Bohm's formulation of quantum mechanics that doesn't involve differentiation or integration of any equations of motion. At each time, $t=n\delta t$ ($n=1,2,3\dots$), $N$ particle positions are randomly sampled from the quantum probability density. Trajectories are built from the sorted $N$ sampled positions at each time. These trajectories become the exact Bohm solutions in the limits $N\rightarrow\infty$ and $\delta t\rightarrow 0$. Higher dimensional problems can be solved by this method for separable wave functions. Several examples are given, including the two-slit experiment. 
\end{abstract}


\maketitle

\section{\label{intro}Introduction}
Given a solution to Schr\"odinger's wave equation, $\psi(x,t)=R(x,t)e^{i S(x,t)/\hbar}$ for real $R(x,t)$ and $S(x,t)$, Bohm's formulation of quantum mechanics \cite{Bohm1952, Bohm1993, Holland1993} defines the equation of motion for particle trajectories in one-dimension as,
\begin{equation}
\label{qNewton2}
	m\ddot x =-\frac{\partial V}{\partial x} - \frac{\partial Q}{\partial x}, 
\end{equation}
where $V$ is the classical potential, and the quantum potential $Q=-(\hbar^2/2 m R)\partial^2 R/\partial x^2$. An ensemble of particle trajectories will maintain a density that is equal to the quantum mechanical probability density $\rho(x,t)=|\psi(x,t)|^2$ if the particle velocities are further constrained to obey the so called {\it guidance law},
\begin{equation}
\label{guidancelaw}
	\dot x=\frac{1}{m}\frac{\partial S}{\partial x}\bigg|_{x=x(t)}.
\end{equation}
Even though the quantum potential does have some explanatory usefulness, typically if $\psi(x,t)$ is known, one only needs to solve the guidance law to obtain the particle trajectories. The guidance law, in general, is impossible to solve in closed form and must be solved numerically \cite{Holland1993, Wyatt2005}.

In Brandt et al. \cite{Brandt1998, Brandt2001} it was shown that the Bohm trajectories can be defined by a concept they have named {\it quantile motion} (a purely non-physical probability concept). A unique trajectory is obtained by requiring that the cumulative probability function (CPF) for the quantum probability density $\rho(x,t)=|\psi(x,t)|^2$ be constant along the trajectory,
\begin{equation}
\label{cpfeq}
	CPF(x)=P=\int_{-\infty}^{x_P(t)} \rho(x,t)\; dx
	={\rm constant},
\end{equation}
where the trajectory is given by $x_P(t)=CPF^{-1}(P={\rm constant})$. Alternatively, one could think of the constant CPF restriction as total left (or right) probability conservation. Like Bohm's equations of motion, the quantile motion solutions can not be obtained in closed form (except for particular probability densities), and therefore, again, must be solved numerically. 

In this paper, we present a Monte Carlo method for a given one-dimensional probability density, that approximates Bohm trajectories \textit{without solving} either of Bohm's equations of motion [Eqs. (\ref{qNewton2}, \ref{guidancelaw})], or the inverse of the CPF in  (\ref{cpfeq}). In other words, our method does {\it not} need to take a derivative or an integral at any point in the process. This fact demonstrates clear computational advantage over the the typical numerical techniques used to solve Eqs. (\ref{guidancelaw}) and (\ref{cpfeq}). The method takes two parameters: the number of particles in the ensemble $N$, and the size of each discrete time step $\delta t$. It is shown that the approximate trajectories become the Bohm trajectories exactly in the limits $N\rightarrow \infty$ and $\delta t\rightarrow 0$. A possible extension of the method into higher dimensions is discussed, and it is found that the method can be used to find Bohm trajectories for those cases of a separable wave function. Finally, we demonstrate the method for the harmonic oscillator, the free particle, and the two-slit experiment. Also provided is a two-dimensional infinite square well example for a separable wave function.

\section{\label{dsmethod}The Density Sampling Method}

\subsection{Description}
\begin{figure}
\begin{center}
\includegraphics[width=3in]{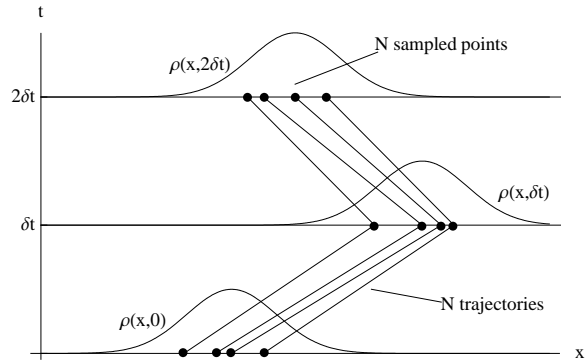}
\caption{\label{dsmfigure}The Density Sampling Method. At each time $t=n\delta t$ ($n=1,2,3,\dots$), $N$ points are sampled from the probability density $\rho(x,t)$ and sorted. Trajectories are constructed by joining the $i$-th sorted point from each time step.}
\end{center}
\end{figure}

\begin{figure}
\begin{center}
\includegraphics[width=3in]{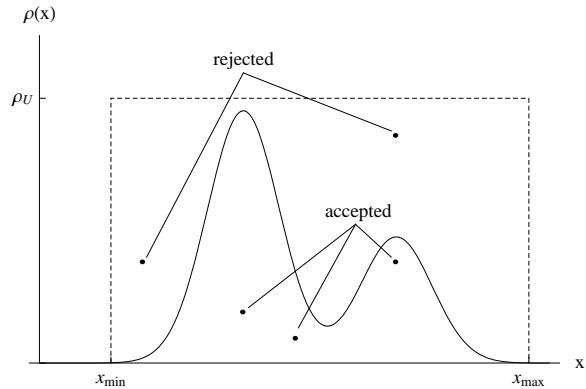}
\caption{\label{vNfigure}The von Neumann Acceptance-Rejection Method. For a given density $\rho(x)$, a set of $x$-points is generated by uniformly placing random dots on the graph. If a dot is under the density curve, that dot's $x$ value is place in the set.}
\end{center}
\end{figure}

The method assumes that a probability density, $\rho(x,t)$, is known and given. The evolution of the density is depicted by an ensemble of $N$ particles, see \fref{dsmfigure}. The trajectories of the $N$ particles are constructed from locations at times $t=n\delta t\  (n=1,2,3,\dots)$ with step size $\delta t$. At each time, the probability density is {\it sampled} to generate a set of $N$ possible $x$-points. The $x$-points are sorted numerically. The $i$-th trajectory in the ensemble is built from the $i$-th $x$-point of the sorted $N$ points at each time step. Though there are many ways to generate a set of $N$ points sampled from a given distribution \cite{randomnumber1, randomnumber2}, in the examples below (\sref{exam}), we use the von Neumann acceptance-rejection method \cite{Neumann1951}, see \fref{vNfigure}, with a uniform proposal distribution $\rho_U$.

The quality of the trajectories generated relies on how one chooses the number of particles in the ensemble $N$, and on the size of each time step $\delta t$. We place the following restrictions for these two parameters,
\begin{equation}
\label{Ndtconditions}
	N \gg 2 L \rho_{\rm max}
	\qquad{\rm and}\qquad
	N\delta t \approx \frac{L}{\epsilon},
\end{equation}
where $L$ is the width of the domain of the $x$ coordinate (generally this is limited to a range of values where the density is essentially non-zero), $\epsilon$ is a small number $0<\epsilon\ll 1$ with the dimensions of a speed, and $\rho_{\rm max}$ is the maximum value of the density for all positions $x\in L$ and for all times between the initial and final times. 

\subsection{Connection to Bohmian Mechanics}
The method constructs the $i$-th trajectory from the $i$-th $N$ sampled and sorted points at each time step. The particle trajectories, therefore, do not intersect by design (a familiar property of Bohmian trajectories). Hence, between successive time steps the approximate size of the maximum change in position $\delta x$ is of the order $\delta x\approx 2 L/N$. Identifying the density sampled trajectory as $x_{DS}(t)$, and the Bohm trajectory as $x_B(t)$, we assume at $t=0$ that  $x_{DS}(0)=x_B(0)$. We now describe the density sampled trajectory as $x_{DS}(t)=x_B(t)+\delta x(t)$ for some function $\delta x(t)\lesssim 2 L/N$ such that $\delta x(0)=0$.  Computing the cumulative probability function (CPF) value (\ref{cpfeq}) for the density sampled trajectory to first order in $\delta x$,
\begin{equation}
	P_{DS}=\int_{-\infty}^{x_{DS}(t)}\rho(x,t)\;dx
	\approx P_{B} + \rho(x_B(t),t)\delta x(t).
\end{equation}
Recall, the CPF value for the Bohm trajectory $P_B$ is constant. From the restrictions and assumptions above, $\rho_{\rm max}\delta x \ll 1$, therefore, $\rho \delta x \le \rho_{\rm max} \delta x\ll 1$, so the density sampled trajectory fluctuates about the Bohm trajectory.

While the parameter $N$ places the location of the particle close to the actual Bohmian location at each time, the other parameter $\delta t$ (the size of each time step) fixes the density sampled speed approximately equal to the Bohm speed. The difference in the two speeds is on the order of $\delta x/(2\delta t)$, where again $\delta x$ is the size of the fluctuation about the Bohmian trajectory. From  (\ref{Ndtconditions}), we find that $\delta x/(2\delta t)\approx \epsilon\ll 1$, or that the difference in the speeds in quite small. We note that the density sampled trajectory will become the Bohm trajectory for $N\rightarrow \infty$ and $\delta t\rightarrow 0$.

\section{\label{higherdim}Extension to Higher Dimensions}
In general, the one-dimensional density sampling method described above can not be extended into higher dimensions. In higher dimensions there is no natural ordering to sort the $N$ sampled points, and therefore, no way to identify the $i$-th position in the ensemble at each time step like in the one-dimensional case. However, the quantile motion concept (\ref{cpfeq}) can be used independently on each coordinate in higher dimensions when the wave function is separable as shown in \ref{higher}. Since the one-dimensional density sampled trajectory fluctuates about the Bohm trajectory defined by  (\ref{cpfeq}), the one-dimensional density sampling method can be used independently on each coordinate to generate higher dimensional Bohm trajectories for those cases of a separable wave function. We also note that the method will generate higher dimensional Bohm trajectories for wave functions that are nearly separable \cite{Poirier1997}, by applying the method independently on each coordinate of the separable part of the wave function.

\section{\label{exam}Examples}
In the examples below, the probability density is determined in the usual way $\rho(x,t)=|\psi(x,t)|^2$. The wave function was chosen to be non-stationary so that $\partial \rho/\partial t\ne 0$. For comparison, the Bohmian trajectories are computed from  (\ref{guidancelaw}) or its equivalent,
\begin{equation}
	\dot x = \frac{\hbar}{2 m i}
	\left(\frac{\psi^\ast \partial \psi/\partial x
	- \psi \partial \psi^\ast/\partial x}{\rho}\right).
\end{equation}
In all cases, the range $L$ of possible position values $x$ was limited to an area where the probability density was essentially non-zero. From  (\ref{Ndtconditions}), the number of particles in the ensemble (or the number of sampled points) $N$ was approximately equal to $2 L \rho_{\rm max}\times 10^3$, while $\epsilon$ was taken to have a numerical value of the order $10^{-3}$.

\subsection{\label{harm}Harmonic Oscillator}
\begin{figure}
\begin{center}
\includegraphics[width=3in]{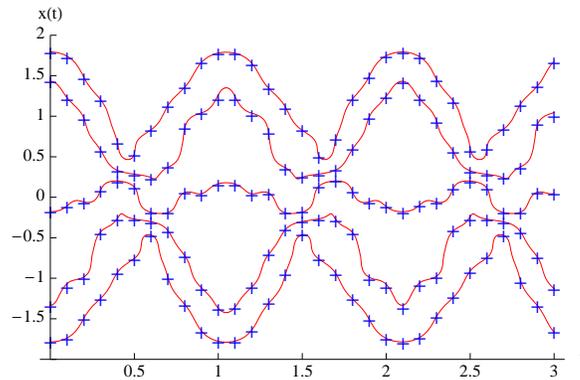}
\caption{\label{harmfigure}(color online). Harmonic oscillator with wave function as a superposition of ground and the first three odd excited states. The Bohmian trajectories are solid lines while the density sampled trajectories are plotted as plus ($+$) signs. The plot is in naturalized units.}
\end{center}
\end{figure}

The harmonic oscillator wave function was a superposition of the ground state and the first three odd excited states,
\begin{equation}
	\psi(x,t)=\frac{1}{2\sqrt{a\sqrt{\pi}}}
	e^{-\frac{x^2}{2a^2}}
	\sum_{n=0,1,3,5}\frac{H_n(x/a)e^{-iE_n t/\hbar}}{\sqrt{n!\ 2^n}}
\end{equation}
where $a=\sqrt{\hbar/m\omega}$, $E_n=\hbar\omega(n+1/2)$, and $H_n$ are the Hermite polynomials. Naturalized units were used so that $\hbar=1$, $\omega=3$, and $m=1$. The range of time was $t\in [0,3]$, and the size of each time step was $\delta t=0.1$. The range of the possible positions was $x\in [-5,5]$ (an area where the density was essentially non-zero). The number of particles in the ensemble was $N=10^4$. Five of the resulting density sampled trajectories ($+$) are shown in \fref{harmfigure} against the actual Bohm trajectories. Notice that the density sampled trajectories are able to depict the oscillatory behavior of the Bohm trajectories rather well.

\subsection{\label{free}Free Particle}

\begin{figure}
\begin{center}
\includegraphics[width=3in]{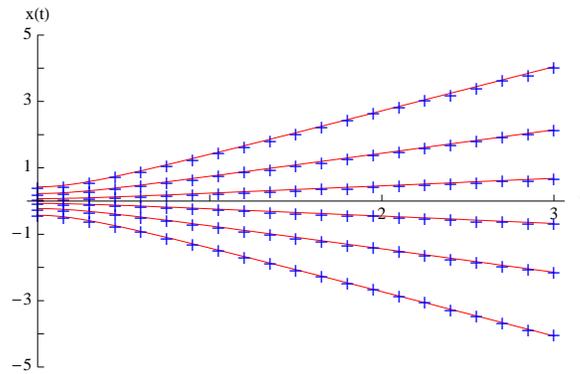}
\caption{\label{freefigure}(color online). Free particle with an initial wave function of a Gaussian centered around zero. The Bohmian trajectories are solid lines, and the density sampled trajectories are plotted as plus ($+$) signs. The plot is in naturalized units. The ensemble of trajectories demonstrates the familiar spreading of the wave packet.}
\end{center}
\end{figure}

The wave function for the free particle (assumed to be Gaussian initially with width $a$) was taken to be,
\begin{equation}
	\psi(x,t)= \left(\frac{2 a}{\pi}\right)^{1/4}
	\frac{e^{-a x^2/[1+(2 i \hbar a t/m)]}}
	{\sqrt{1+(2 i\hbar a t/m)}}
\end{equation}
Naturalized units were used so that $\hbar=1$, $m=1$, and $a=\pi/2$. Again, $t\in [0,3]$ with time steps of $\delta t=0.15$. The number of particles in the ensemble was $N=10^5$. Six of the resulting density sampled trajectories are plotted in \fref{freefigure} superposed on top of their corresponding Bohm trajectory. The trajectories depict the familiar spreading of the wave packet.

\subsection{\label{twoslits}Two-Slit Experiment}
\begin{figure}
\begin{center}
\includegraphics[width=6.5in]{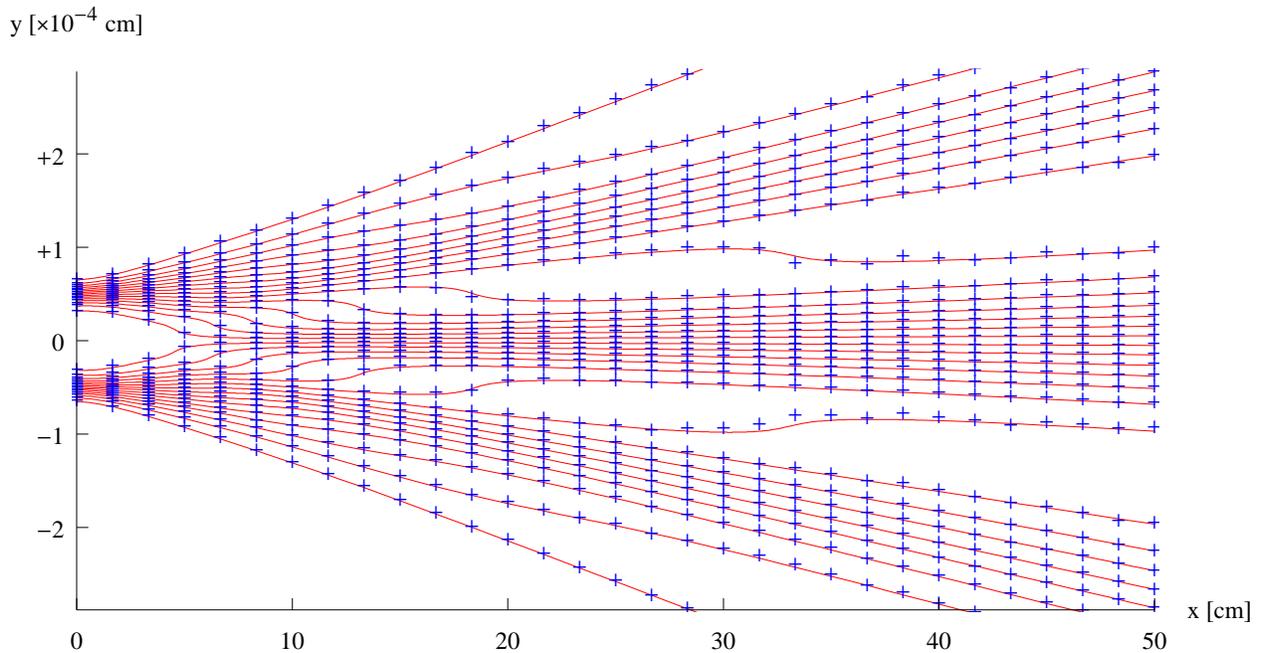}
\caption{\label{twoslitsfigure}(color online). Two-Slit Experiment as described in \S5.1.2 of P.R.~Holland, {\it The Quantum Theory of Motion: An Account of the de Broglie-Bohm Causal Interpretation of Quantum Mechanics}, (Cambridge University Press, New York, 1993). The density sampled trajectories ($+$) are plotted superposed on the Bohmian trajectories (solid). The initial positions are assumed Gaussian in the slits (left side of figure), and the ensemble of trajectories makes the familiar bands of bright and dark on the screen (right side of the figure).}
\end{center}
\end{figure}

This two-slit example is from \S5.1.2 in Holland \cite{Holland1993}. At first this problem seems to be two-dimensional. However, the motion along the coordinate from the slits to the screen [$x$ in \fref{twoslitsfigure}] is assumed uniform, thus the probability density is effectively one-dimensional. To allow for easier computation the experimental values given in Holland's book were rescaled so that $\hbar=1$, $m=1$, and the total time between the slits and screen was $t_{\rm max}=100$. The range of possible positions was $y\in [-129.668,+129.668]$. The number of particles in the ensemble was $N=10^5$, and $\delta t=t_{\rm max}/30$. The resulting trajectories were then rescaled back to Holland's numbers for plotting. Thirty of the density sampled trajectories ($+$) are plotted against their Bohm counterpart in \fref{twoslitsfigure}. The trajectories manifest the familiar bright and dark bands of the two-slit intensity pattern on the screen (left side of figure). Also, notice the size of the  fluctuations about the Bohm trajectory in the different regions. In high density regions the method does better since $\delta x\propto 1/N$ is smaller. But in low density regions (between the bright bands) $\delta x$ is larger due to less particles being there.

\subsection{\label{2dwell}Separable Wave Function in a Two-Dimensional Infinite Square Well}

\begin{figure}
\begin{center}
\includegraphics[width=3in]{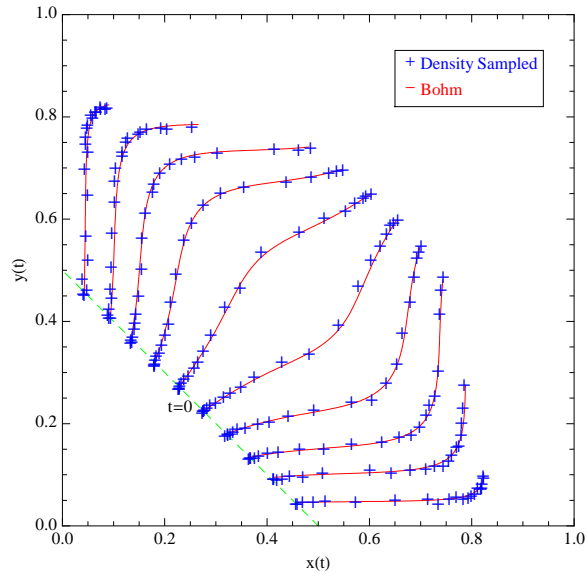}
\caption{\label{2dwellfigure}(color online). Comparison of the density sampled ($+$) and the Bohm trajectories for a separable wave function in the two-dimensional infinite square well of size $1\times 1$ in naturalized units. The initial position of each trajectory lies on the line from $(0.5,0)$ to $(0,0.5)$.}
\end{center}
\end{figure}

In \fref{2dwellfigure} is a comparison of the density sampled trajectories and their Bohm counterpart for the two-dimensional infinite square well. The separable wave function was assumed to be $\psi(x,y,t)=\psi_x(x,t)\psi_y(y,t)$, where,
\begin{equation}
	\psi_x(x,t)=\sqrt{\frac{1}{L}}
	\left(
	\sin\left(\frac{\pi x}{L}\right)e^{-i E_1 t/\hbar}
	+ \sin\left(\frac{2 \pi x}{L}\right)e^{-i 4 E_1 t/\hbar}
	\right),
\end{equation}
and a similar expression for $\psi_y(y,t)$. The energy $E_1=\pi^2 \hbar^2/2 m L^2$, and naturalized units were used so that $m=1$, $\hbar=1$, and the width of the well in each direction taken to be $L=1$. The one-dimensional density sampling method was used independently for each coordinate. The time was $t=n\delta t\in[0,1]$ for $n=1,2,3\dots$, and the size of each time step was $\delta t=0.05$. The number of particles in each coordinate's ensemble was $N=10^4$. Note, in general, a particle's place in the each coordinate's ensemble is not the same. The density sampled trajectory points ($+$) are plotted superposed on top of the corresponding Bohm trajectories. For the separable wave function, the density sampled trajectories are again identical to the Bohm trajectories.

\section{\label{conclusion}Concluding Remarks}
The density sampling method described in this paper shows that one can generate approximate Bohmian trajectories without differentiation or integration of any equations of motion. The method merely evaluates the given probability density at various locations and times. It uses two adjustable parameters. The first being the size of the set of sampled points $N$, and the second the size of each time step $\delta t$. It was shown that as $N\rightarrow\infty$ and $\delta t\rightarrow 0$ the density sampling method trajectories become the Bohm trajectories. Like the quantile motion idea, this method can generate the unique one-dimensional Bohmian trajectories without any additional physical assumptions to standard quantum mechanics. In addition, this method can be used to construct one-dimensional Bohmian trajectories for densities where the wave function is unknown or doesn't exist (i.e., non-quantum problems). Further, if a known higher dimensional wave function is separable (or nearly separable), then the method can be applied independently to each coordinate, thus constructing higher dimensional Bohmian trajectories as well.

\ack
R.E.W. thanks the Robert Welch Foundation for financial support.

\appendix

\section{\label{higher}Extension of Quantile Motion into Higher Dimensions}

The extension of the quantile motion into higher dimensions is also discussed in Brandt et al. \cite{Brandt1998}. They show that instead of the total left (or right) probability being conserved in one-dimension, that in higher dimensions the probability is conserved inside a volume that is enclosed by a surface of Bohmian trajectories. This property, however, is not unique to Bohmian mechanics. Any velocity field $\dot{\bf x}$ will conserve the probability inside the volume in configuration space since \cite{Wyatt2005}(chapter four),
\begin{equation}
	\frac{d\rho}{dt}=-\rho\nabla\cdot \dot{\bf x}
	\quad{\rm and}\quad
	\frac{dJ}{dt}=+J\nabla\cdot \dot{\bf x},
\end{equation}
where $\rho$ is the probability density and $J$ is the Jacobian such that the volume changes as $dV(t)=J dV_0$. The probability inside this evolving volume is,
\begin{equation}
	P_{\rm in}=\int \rho \; dV(t) = \int \rho J \; dV_0. 
\end{equation}
Which implies that $dP_{\rm in}/dt = 0$. Therefore, any velocity field $\dot{\bf x}$ will conserve the total probability inside a volume that in enclosed by a surface of trajectories solved from $\dot{\bf x}$. This is in contrast to what is found in the one-dimensional case, where the velocity field that satisfies the total left (or right) probability conservation is unique.

However, the quantile motion concept can be used to generate trajectories in higher dimensions if one uses the marginal distribution for each coordinate analogously to the cumulative probability distribution (CPF) in one-dimension. Suppose the system can be described by $N$ Cartesian coordinates, then the corresponding definition of ~(\ref{cpfeq}) is,
\begin{equation}
\label{newtotalleft}
	P_i = \int_{-\infty}^{x_i(t)}\rho_i(x_i,t)\;dx_i
	\qquad{\rm for}\quad{i=1,2,\dots,N},
\end{equation}
where $\rho_i(x_i,t)$ is the marginal distribution for the $i$-th coordinate. Again, we assume that the particles are conserved so,
\begin{equation}
	\frac{\partial \rho}{\partial t}
	+\sum_{i=1}^N \frac{\partial}{\partial x_i}(\rho \dot x_i)
	=0.
\end{equation}
Partially integrating this continuity equation we have that,
\begin{eqnarray}
	&&\int_{-\infty}^{+\infty}..\int_{-\infty}^{x_i(t)}..\int_{-\infty}^{+\infty}\frac{\partial \rho}{\partial t}\;dx_1\dots dx_N\nonumber
\\
	&&+\sum_{j=1}^N \int_{-\infty}^{+\infty}..\int_{-\infty}^{x_i(t)}..\int_{-\infty}^{+\infty}\frac{\partial (\rho \dot x_j)}{\partial x_j}\;dx_1\dots dx_N
	=0.
\end{eqnarray}
Interchanging the partial derivative with respect to time and performing the $\pm \infty$ integrations, and assuming that the density is zero at $\pm \infty$, only the $j=i$ integrals on the second term survive,
\begin{eqnarray}
	&&\int_{-\infty}^{x_i(t)}\frac{\partial \rho_i}{\partial t}\;dx_i\nonumber
	\\
	&&\qquad+\int_{-\infty}^{+\infty}..\int_{-\infty}^{+\infty}\rho \dot x_i \;dx_1\dots dx_{\neq i}\dots dx_N 
	=0.
\end{eqnarray}
Then with the expression above, and differentiating ~(\ref{newtotalleft}) with respect to time,
\begin{equation}
	\frac{dP_i}{dt}= \rho_i \dot x_i
	-\int_{-\infty}^{+\infty}..\int_{-\infty}^{+\infty}\rho \dot x_i \;dx_1\dots dx_{\neq i}\dots dx_N.
\end{equation}
Hence, the $i$-th coordinate, in general, doesn't conserve total left probability of the marginal distribution since $\dot x_i$ could depend on the other coordinates. Suppose, however, that $\dot x_i=\dot x_i (x_i,t)$ (i.e. the motion along the $i$-th coordinate is independent), then $dP_i/dt=0$, and the total left probability is conserved for the marginal distribution $\rho_i$. 

In higher dimensional Bohmian problems the guidance law (\ref{guidancelaw}) for the $i$-th coordinate becomes,
\begin{equation}
	\dot x_i = \frac{1}{m}\frac{\partial S(x_1,x_2,\dots,x_N;t)}{\partial x_i}.
\end{equation}
If the wave function is separable, then,
\begin{equation}
	\psi=\psi_1(x_1,t)\psi_2(x_2,t)\cdots\psi_N(x_N,t).
\end{equation}
The probability density is also separable, $\rho=\rho_1(x_1,t)\rho_2(x_2,t) \dots\rho_N(x_N,t)$, and the phase becomes $S=S_1(x_1,t) + S_2(x_2,t)+\dots+S_N(x_N,t)$. Which implies that $\dot x_i = (1/m)\partial S_i(x_i,t)/\partial x_i$, or that the velocity field for the $i$-th coordinate is independent of the other coordinates, for all $i=1,2,\dots,N$. Hence, one can use the one-dimensional total left (or right) probability conservation, independently for each coordinate, to generate higher-dimensional Bohmian trajectories for a separable wave function.


\section*{References}

\begin{thebibliography}{00}
\bibitem{Bohm1952}
Bohm~D 1952 {\it Phys. Rev.} {\bf 85} 166, 180

\bibitem{Bohm1993}
Bohm~D and Hiley~B~J 1993  {\it The Undivided Universe} (New York:Routledge)

\bibitem{Holland1993}
Holland~P~R 1993 {\it The Quantum Theory of Motion: An Account of the de Broglie-Bohm Causal Interpretation of Quantum Mechanics} (New York: Cambridge University Press)

\bibitem{Wyatt2005}
Wyatt~R~E 2005 {\it Quantum Dynamics with Trajectories: Introduction to Quantum Hydrodynamics} (New York: Springer)

\bibitem{Brandt1998}
Brandt~S, Dahmen~H, Gjonaj~E and Stroh~T 1998 {\it Phys.
  Lett.} A {\bf 249} 265

\bibitem{Brandt2001}
Brandt~S and Dahmen~H~D 2001 {\it The Picture Book of Quantum Mechanics} (New York: Springer-Verlag) 3rd ed

\bibitem{randomnumber1}
Gentle~J~E 2003 {\it Random Number Generation and Monte Carlo Methods} (New York: Springer-Verlag) 2nd ed

\bibitem{randomnumber2}
Lewis~T~G 1975 {\it Distribution Sampling for Computer Simulation} (Massachusetts: Lexington Books)

\bibitem{Neumann1951}
von~Neumann~J 1951 {\it Nat. Bur. Standards} {\bf 12} 36

\bibitem{Poirier1997}
Poirier~B 1997 {\it Phys. Rev.}~A {\bf 56} 120
 
\hskip-1em Colavecchia~F~D, Gasaneo~G and Garibotti~C~R 1998 {\it Phys. Rev.}~A {\bf 57} 1018

\end{thebibliography}
\end{document}